\documentclass[12pt]{iopart}

\usepackage{iopams}  
\usepackage{graphicx}
\usepackage[breaklinks]{hyperref}

\usepackage{graphicx}

\expandafter\let\csname equation*\endcsname\relax

\expandafter\let\csname endequation*\endcsname\relax
\usepackage{amsmath}

\newcommand{\ket}[1]{|{#1}\rangle}

\begin{document}

\title[Polarization entanglement and quantum beats of photon pairs
\ldots]{Polarization entanglement and quantum beats of photon pairs from
  four-wave mixing in a cold $^{87}$Rb ensemble}

\author{Gurpreet~Kaur~Gulati}
\address{Center~for~Quantum~Technologies,
  National~University~of~Singapore,
  3~Science~Drive~2,
  Singapore, 117543}
\author{Bharath~Srivathsan}
\address{Center~for~Quantum~Technologies,
  National~University~of~Singapore,
  3~Science~Drive~2,
  Singapore, 117543}
\author{Brenda~Chng}
\address{Center~for~Quantum~Technologies,
  National~University~of~Singapore,
  3~Science~Drive~2,
  Singapore, 117543}
\author{Alessandro~Cer\`{e}}
\address{Center~for~Quantum~Technologies,
  National~University~of~Singapore,
  3~Science~Drive~2,
  Singapore, 117543}
\author{Christian Kurtsiefer$^{1,2}$}
\address{$^1$Center~for~Quantum~Technologies,
  National~University~of~Singapore,
  3~Science~Drive~2,
  Singapore, 117543}
\address{$^2$Department~of~Physics,
  National~University~of~Singapore,
  2~Science~Drive~3, 
  Singapore, 117542}
\ead{christian.kurtsiefer@gmail.com}
\date{\today}

\begin{abstract}
  We characterize correlations in polarization and time of photon pairs 
  generated from a cold cloud of $^{87}${Rb} atoms via a four-wave mixing
  process in a cascade level scheme. 
  Quantum state tomography reveals entangled polarization states of high purity
  for each of the decay paths through two different intermediate hyperfine
  levels.
  When allowing both decay paths, we observe quantum beats in time-resolved
  correlation measurements.
\end{abstract}

\pacs{
42.50.Dv, 	
03.65.Wj, 	
03.67.Bg 	
}

\maketitle
\section{Introduction}
Time-correlated and entangled photon pairs have been an important resource for a wide range
of quantum optics experiments, ranging from fundamental tests
\cite{Brunner:2014}
to applications in quantum communication,
cryptography,
teleportation
and computation~\cite{Nielsen:2004}.

The first sources of correlated photon pairs were based on a cascade decay in neutral atoms~\cite{Fry:1973bq,aspect:82}.
The cascade imposes a time correlation, and with an appropriate choice of the
geometry and intermediate states it is possible to observe a strong
non-classical correlation in the polarization of the photons.
These sources are rarely used today in quantum optics experiments for their
limited generation rates, and have been superseded by schemes based on
three-wave mixing in non-linear  optical crystals, or four-wave mixing (FWM)
in optical fibers \cite{fiorentino2002all}. To obtain a narrow 
optical bandwidth, it is possible to use near-resonant transitions in
atoms to provide the large third order nonlinear susceptibility
for efficient FWM, leading to photon pairs with a central wavelength
matching those of the transitions
involved~\cite{Kuzmich:2003,Balic:2005jg,Chaneliere:2006}.
These sources are a hybrid between the traditional atomic cascade approaches,
and those based on three- or four-wave mixing in solids in the sense that they
deliver both a useful pair rate collected into single mode optical fibers, and
exhibit interesting temporal correlations.

In this paper, we present the characterization of a source of time-correlated
and polarization-entangled photon pairs based on four-wave mixing in a cold
cloud of $^{87}${Rb} atoms. The involved atomic levels,
selected by the choice of pump and target wavelengths,
form a cascade decay scheme, providing an asymmetrical time correlation
similar to the one from the cascade decay of single atoms.

\section{Polarization entanglement}
Entanglement between photons can be established in several degrees of freedom
like time bins, polarization, and orbital angular momentum \cite{RevModPhys.74.145},
with polarization entanglement having been extensively studied due to its
robustness and the availability of very stable optical elements for
manipulating and detecting the polarization of single photons~\cite{Englert:01}. 

Observation  of photon polarization entanglement begin with 
early photon pair sources based on cascade decays in atomic
beams~\cite{Aspect:1981}, followed by
spontaneous parametric down conversion in nonlinear optical
crystals~\cite{Kwait:1995}, cold~\cite{Chaneliere:2006} and
hot~\cite{Willis:2011} atomic vapors, and recently also in cascade emission
from quantum dots~\cite{muller:2014}.
In this paper, we characterize the correlation polarization properties of
nearly Fourier-limited photon pairs generated from the cold cloud of
$^{87}${Rb}~\cite{Srivathsan:2013, Gulati:2014}.

\section{Experimental setup}
\begin{figure}
\hfill\includegraphics[width=0.85\columnwidth]{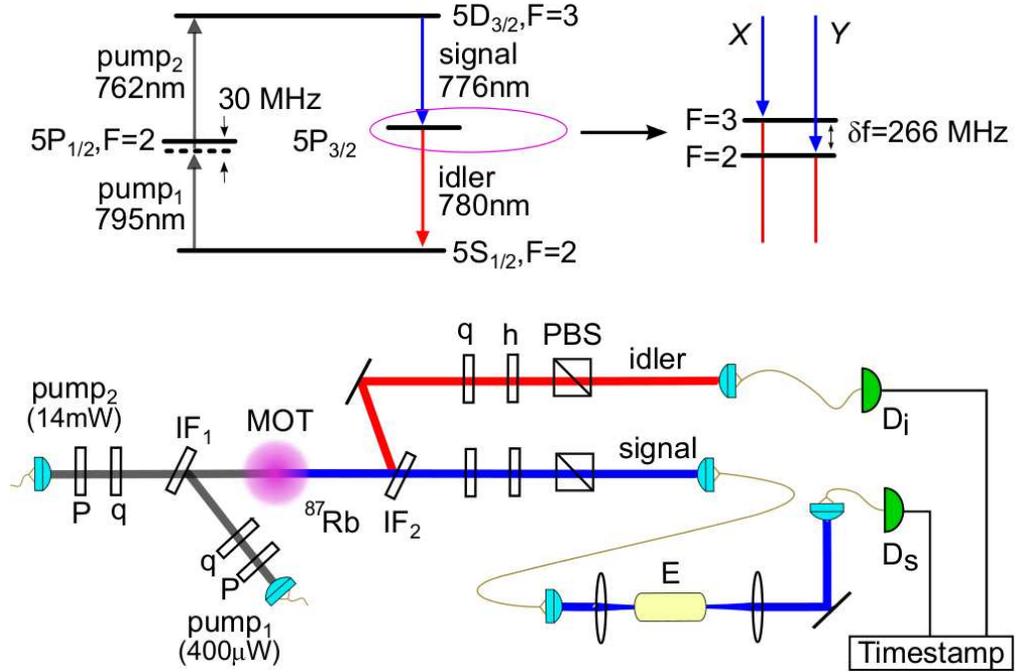}
    \caption{\label{fig:schematic} Level scheme for four-wave mixing in
      $^{87}$Rb, and schematic of the experiment. A first
      interference filter (IF$_1$) combines the two pump beams in a
      co-propagating geometry inside the cloud, a second one (IF$_2$)
      separates the signal and idler photons from residual pump light. The
      pump polarizations can be freely chosen with polarizers (P) and quarter
      wave plates (q). A stack of quarter wave plate, half wave
      plate (h), and polarizing beam splitter (PBS) in each collection mode
      can select any arbitrary polarization. A solid etalon (E) can be used to
      select light from only one for the decay paths $X$ and $Y$. Di, Ds:
      Avalanche Photodetectors.
    }
\end{figure}
The experimental setup is similar to our earlier work~\cite{Srivathsan:2013,
  Gulati:2014}, but uses a collinear beam geometry (see 
figure~\ref{fig:schematic}).
An ensemble of  cold $^{87}${Rb} atoms is prepared with a magneto-optical trap
(MOT) of
optical density OD\,$\approx 32$ for light resonant to the 
$5$S$_{1/2},\,F=2$ $\rightarrow$ $5$P$_{3/2},\,F=3$ transition.
The atoms are excited from $5$S$_{1/2},F=2$ to $5$D$_ {3/2},\,F=3$ via a
two-photon transition, with a two-photon detuning of $\approx5$\,MHz.
Pump beams of wavelength 795\,nm and 762\,nm overlap in a  co-propagating geometry inside the cloud.
The 795\,nm pump is red detuned by 30\,MHz from the intermediate level
$5$P$_{1/2},\,F=2$ to reduce the incoherent scattering rate.
From  the $5$D$_ {3/2},\,F=3$ excited level, atoms can decay through
several paths. We select ``signal'' photons around 776\,nm, and ``idler''
photons around 780\,nm with interference filters of 3\,nm FWHM bandwidth.
Within this bandwidth, two
decays can be observed
(figure~\ref{fig:schematic}, top right):
decay $X$ through the hyperfine level $5$P$_ {3/2},\,F=3$, and $Y$ going through $5$P$_ {3/2},\,F=2$.

Energy conservation and phase matching results in the generation of signal
and idler photon pairs from both decay paths with a frequency
difference of  $\delta$=266\,MHz corresponding to the hyperfine splitting
of the intermediate level.
The generated photons are collected into single-mode fibers  and detected by
avalanche photodetectors (quantum efficiency $\approx$\,40\%, jitter time
$\approx$\,1\,ns).
In the experiment, we cycle between a 150\,$\mu$s long cooling period with
the MOT turned on, and a 10\,$\mu$s long period for pair generation.

\section{Polarization state tomography}
\begin{figure}
  \begin{center}
    \hfill\includegraphics[width=0.85\columnwidth]{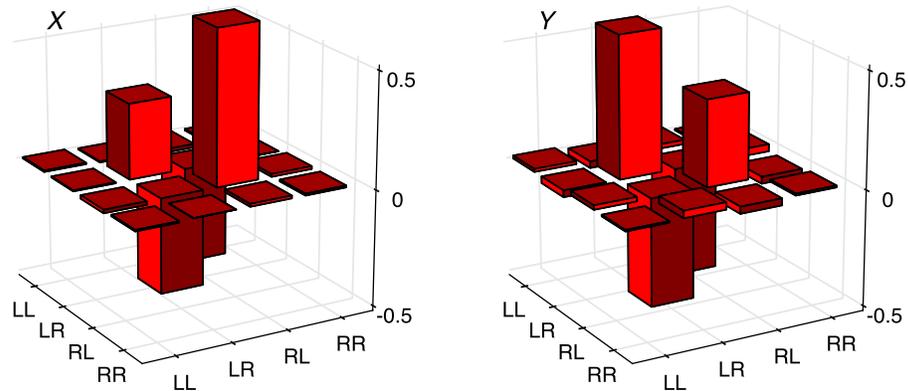}
    \caption{\label{fig:tomo} 
    Tomographic reconstruction of the polarization state $\rho$  (real part)
    for 
    biphotons generated via
    decay path $X$ (on the left), and $Y$ (right),
    for pump modes set to orthogonal
    circular polarizations. The imaginary part of all matrix elements is below
    0.09 and is not shown.}  
    \end{center}
\end{figure}
We investigate the polarization state of the photon pairs for 
decay paths $X$ and $Y$ independently.
We select light in the signal mode (see figure~\ref{fig:schematic}) using 
a 2\,cm long solid fused silica etalon with a transmission bandwidth of 52\,MHz
(FWHM).
The etalon is temperature-tuned to match to the resonance frequency of either
the 
$5$P$_{3/2},\,F=3\,\rightarrow\,5$D$_{3/2},\,F=3$ or
$5$P$_{3/2},\,F=2\,\rightarrow\,5$D$_{3/2},\,F=3$ transition; its  
temperature is stabilized to within 1\,mK  to minimize frequency drifts. 

We completely characterize the polarization state of photon pairs
via quantum state
tomography~\cite{James:2001} by projective detections of individual
photons in a combination of linear and circular polarizations. For this, we
insert quarter and half wave plates (q, h) followed by beam splitter
cubes (PBS) in the signal and idler modes.
Figure~\ref{fig:tomo} shows the real part of the reconstructed
biphoton states  $\rho_{X}$ and $\rho_{Y}$ corresponding to decay paths
$X$ and $Y$. The imaginary parts of all  elements are
smaller than~$\pm0.09i$. The strong off-diagonal elements ($LR$, $RL$) signify
the polarization entanglement. 
From the reconstructed matrices, we can extract typical entanglement measures 
for the state; we evaluate the concurrence $C$~\cite{Coffman:2000}, and the
entanglement of formation $E$~\cite{Wootter:1998}. Furthermore, we can also
determine the purity $P=\mathrm{Tr}[\rho_{X,Y}^2]$ of the biphoton state for
each decay path. The values of these indicators are given in
table~\ref{tab:pol}.

\begin{table}
  \begin{center}
    \begin{tabular}{lp{1em}cc}
      \hline\hline
       && $X$ & $Y$\\ \hline
      Purity $P$ && 0.92$\pm$0.02   & 0.96$\pm$0.03 \\
      Concurrence $C$ && 0.89$\pm$0.01   & 0.94$\pm$0.01 \\
      Entanglement of formation $E$ && 0.85$\pm$0.03   & 0.98$\pm$0.01 \\ \hline\hline
    \end {tabular}   
    \caption{\label{tab:pol}
    Entanglement indicators for reconstructed states $\rho_{X,Y}$. 
    The uncertainties reflect propagated Poissonian counting statistics of
    contributing coincidence events.}  
  \end{center}
\end {table}

Despite the fact that the atomic ensemble is not prepared in a particular
Zeeman sublevel, the polarization states for photon pairs from both decay
paths show a remarkably high purity.  This is compatible with theoretical
models presented in~\cite{Jenkins:07,stewart_2012}, which we briefly summarize
here.

In a cascade decay,
polarization entanglement arises from indistinguishable decay paths, in our case provided by sufficiently degenerate Zeeman states of each hyperfine level.
With the quantization axis along the beam propagation direction of all modes, we only drive transitions with $\Delta m_F=\pm1$ with
orthogonally circularly polarized pump beams.
In parametric processes~\cite{Boyd2008},
the quantum state of the medium remains
unchanged through the interaction~\cite{D.N_2005}.
Further, rotational symmetry of the atomic cloud in beam propagation direction
implies angular momentum conservation.
Along with the angular momentum selection rules, this
limits the possible polarizations of the generated signal-idler photon pairs
to $\ket{LR}$ and $\ket{RL}$. Since the process is coherent and
$\ket{LR}$ and $\ket{RL}$ are indistinguishable otherwise, the 
resulting state $\ket{\psi}$ of a target mode photon pair can be written as
\begin{equation}\label{eq:cg}
  \ket{\psi}=a_{0}\ket{LR}\,+\,\exp(i\phi_{0})\,a_{1}\ket{RL}\,.
\end{equation}
The probability amplitudes $a_{0,1}$ and the phase $\phi_{0}$ can be 
derived using a model based on the
relative transition strength between different Zeeman
sublevels~\cite{Jenkins:07,stewart_2012} as
\begin{equation}\label{eq:poltomo}
  a_{0,1} = \frac{x_{\alpha_S, \alpha_I}}{\sqrt{\sum\limits_{\alpha_S,\alpha_I=
    \pm 1}(x_{\alpha_S, \alpha_I})^2}}\,,
\end{equation}
where  $\alpha_{S,I}$ are the helicities of the signal and idler photons,
and  $x_{\alpha_S, \alpha_I}$ is the product of relevant Clebsh-Gordan
coefficients~\cite{metcalf_1999}
that couple the individual $\ket{m_F}$ states of 
the different hyperfine levels involved in the four-wave mixing process, and
\begin{eqnarray}
  x_{\alpha_S, \alpha_I}=
  \sum\limits_{m_F=-F_g}^{F_g}
  C_{m_F,-1, m_F-1}^{F_g, 1, F_b}\,C_{m_F-1, 1, m_F}^{F_b, 1, F_e}\,
  C_{m_F-\alpha_S, \alpha_S, m_F}^{F_d, 1, F_e}\,C_{m_F,-\alpha_I, m_F-\alpha_I}^{F_g, 1, F_d}\,,
\end{eqnarray}
where $F_{g,b,e,d}=2, 2, 3, 3$ corresponding to the respective total angular
momentum $F$ of the participating levels. 
From (\ref{eq:poltomo}), we obtain the expected state
$\ket{\psi_X}\approx 0.55\ket{LR}-0.83\ket{RL}$ for the decay path $X$.
The reconstructed state $\rho_X$ matches the expected one with a fidelity of
94$\pm$1\%. For the decay path $Y$, the model predicts state
$\ket{\psi_Y}\approx$ $0.92\ket{LR}-0.39\ket{RL}$, which agrees with $\rho_Y$
with a fidelity of 93$\pm$1\%. 

\section{Transition strength of different decay paths}
\begin{figure}
  \begin{center}
    \hfill\includegraphics[width=0.85\columnwidth]{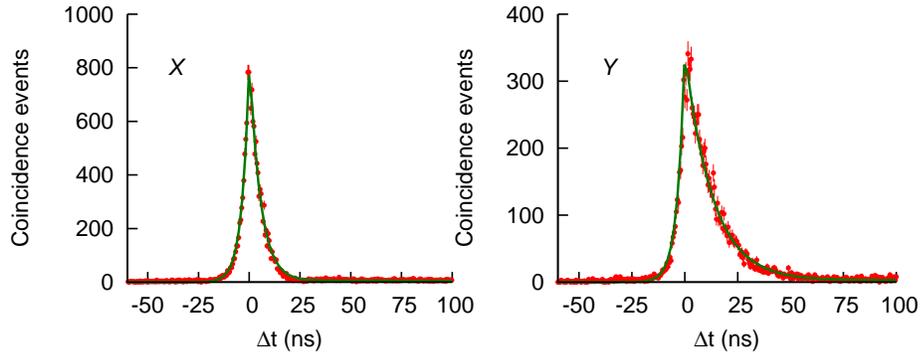} 
    \caption{\label{fig:decaytime}  Coincidence events as a function of
      the detection time difference $\Delta t$ between signal
      and idler photon detection for the decay path $X$  and $Y$,  selected
      with a  temperature tuned etalon.  Integration times were 7 and 14
      minutes, respectively.} 
   \end{center}
\end{figure}
Both decay paths exhibit different decay time constants due to different
transition strengths for the decays. 
The transition $5$P$_{3/2},\,F=3$ to $5$S$_{1/2},\,F=2$ is 2.8
times stronger than from $5$P$_{3/2},\,F=2$~\cite{Steck_2010}.
This results in a higher optical density (OD) for the $\,F=3$
transition~\cite{foot:2005}, and should lead to a faster decay via path $X$~\cite{Jen:2012fq}.
To experimentally investigate this, 
we perform separate time correlation measurements between the detection of signal
and idler photons for each decay path.
The histogram of coincidence events as a function of time delay $\Delta t$
between the detection of signal and idler photons sampled into 1\,ns wide time bins is shown in figure~\ref{fig:decaytime}.
The solid line in both cases shows a fit to a heuristic model, inspired by
the joint effect of a decay time corresponding to the two-photon transition~\cite{Srivathsan:2013,Jen:2012},
and the effect of a finite ring-down time of the filter etalon:
\begin{equation}\label{eq:decaytime}
  G^{(2)}_{X,Y}(\Delta t)= 
  \begin{cases}
    G_0\,\exp(\Delta t/\tau_r) & \mathrm{ for }\; \Delta t<0\\
    G_0\,\exp(-\Delta t/\tau_{X,Y}) & \mathrm{ for }\; \Delta t\geq0\,.\\
  \end{cases}
\end{equation}
For path $X$, we obtain a decay constant $\tau_{X}$=5.6$\pm 0.1$\,ns for an idler photon heralded by a signal photon. 
In the same way, with the etalon tuned to transmit the resonance frequency of
the $5$P$_{3/2},\,F=2$$\rightarrow$  $5$D$_{3/2},\,F=3$ transition for path $Y$,
we find $\tau_{Y}$= 13.1$\pm 0.2$\,ns.
Both decay constants $\tau_{X,Y}$ are shorter than the spontaneous decay time
$\tau_{sp}\,=\, 27$\,ns of the $5$P$_{3/2}$ level of a single Rubidium atom
in free space due to the collective enhancement
effects observed in an optically thick atomic ensemble .

The rise time $\tau_{r}\,=\,3.1\pm0.3$\,ns for decay path $X$
and $\tau_{r}\,=\,3.3\pm0.4$\,ns for  $Y$ is
a consequence of the finite response time of the etalon. 
Both values are compatible with the value of $3.0\pm0.1$\,ns obtained in an
independent characterization of the etalon.

\section{Quantum beats}
\begin{figure}
  \begin{center}
    \hfill\includegraphics[width=0.85\columnwidth]{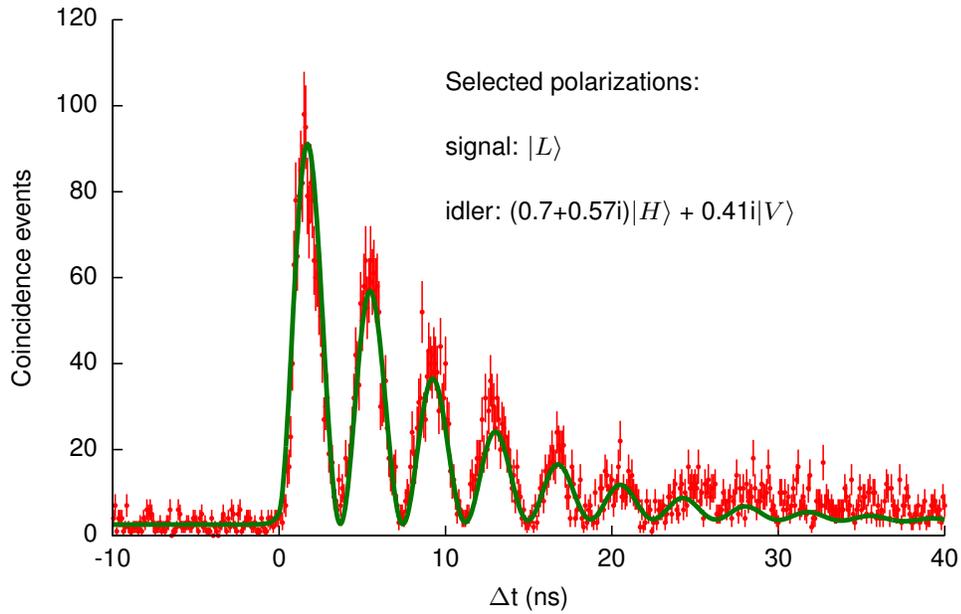}
    \caption{\label{fig:interference_beats}
      Coincidences as a function of the time delay between the detection of signal and idler photons, with no
      etalon in the signal mode.
      Pump 1 and 2 are set to orthogonal linear polarization
      $H$ and $V$, respectively, and signal and idler are projected onto the polarization states $\ket{L}$ and $(0.7+0.57i)\ket{H}+0.41i\,\ket{V}$.
      The observed modulation (``quantum beat'') is associated with the hyperfine splitting of 266\,MHz between $F=3$
      and $F=2$ of the $5P_{3/2}$ level.
      To resolve the oscillations with high contrast, avalanche photodetectors with a low time jitter ($\approx$40\,ps)
      were used for this measurement.
      Due to the lower quantum efficiency of these detectors,
      the total acquisition time is 5 hours.
      }
    \end{center}
\end{figure}

Without the etalon (see figure~\ref{fig:schematic}), the decay paths $X$ and
$Y$ cannot be distinguished by wideband photodetectors. Consequently,
the energy difference between photons from the two paths leads
to a modulation of the time correlation function 
between signal and idler photodetection,
as shown in Figure~\ref{fig:interference_beats}.
This, and other similar phenomena, is known as quantum beats: it was predicted at an early stage
of quantum physics~\cite{breit_1933}, 
and first experimentally observed in pulsed optical excitation of atoms with two
upper states decaying to the same ground state~\cite{Aleksandrov_1964, dodd_1967}.
Quantum beats have also been observed in cascade decay systems of dilute atomic
vapours~\cite{Aspect:1984}, dense thermal atomic
vapours~\cite{Becerra:2011}, and for single ions~\cite{Schug:2014}.

\begin{figure}
  \begin{center}
    \hfill\includegraphics[width=0.85\columnwidth]{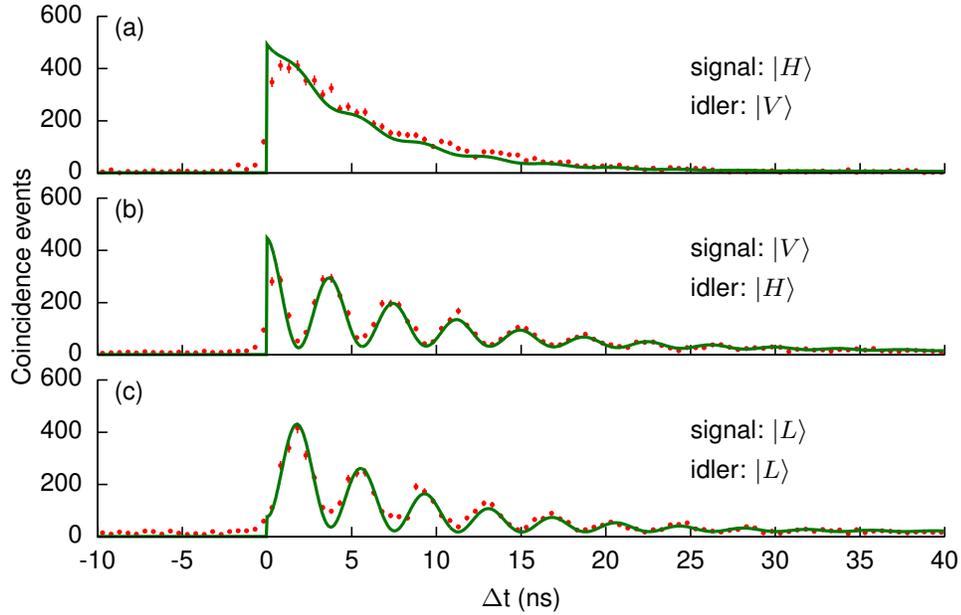}
    \caption{
      \label{fig:beats_phase}
      Coincidence rate as a function of time delay between the detection of signal and idler photons for different choices of polarization of signal and idler photons.
      (a)  The beats are damped by choosing the appropriate polarizations due to suppression of coincidences from decay path $Y$. $R = 2.86*10^{-2}$ and $\phi=\pi$, total acquisition time 9 minutes.
      (b) and (c): different polarization projections change  the phase of the oscillation.
      In this two cases, the relative phase difference is $\pi$. (b) $R = 1.43$ and $\phi=0$, total acquisition time 35 minutes; (c) $R = 0.5$ and $\phi=\pi$, total acquisition time 28 minutes.
      }
  \end{center}
\end{figure}

In our case, the beat frequency $\delta/2\pi$=266\,MHz
is equal to the energy difference between the hyperfine levels
$5$P$_{3/2},\,F=3$ and $F=2$.
The measurements shown in Figure~\ref{fig:interference_beats} are performed with
polarization of pump 1 and 2 set to $H$ and $V$, respectively, 
and  polarizations of signal and idler modes are set to observe
a large interference contrast.

To model the interference between the different decay paths,
we take into account the different coherence time of the emitted photon pairs,
characterized by the time constants $\tau_X$ and $\tau_Y$,
and the relative difference in the generation amplitude $R$, and phase $\phi$.
These last two terms can be calculated via (\ref{eq:poltomo}). 
We express the probability amplitudes $c_{X,Y}$ for paths $X$ and $Y$
as function of the detection time difference $\Delta t$ between
signal and idler photons,
\begin{equation}
  c_X(\Delta t) =\Theta(\Delta t) G_0\, e^{-\frac{\Delta t}{2\tau_X} -i\omega_i \Delta t}\,,\quad\text{and}\quad
  c_Y(\Delta t) =\Theta(\Delta t) G_0R \, e^{-\frac{\Delta t}{2\tau_Y}
    -i(\omega_i+\delta)\Delta t+\phi}\,,
\end{equation}
which interfere to a joint detection probability 
\begin{eqnarray}\label{eq:model}
  G^{(2)} (\Delta t) &=& \left|c_X+c_Y\right|^2\nonumber \\
    &=& \Theta(\Delta t)G_0^2 \left[e^{-\frac{\Delta
          t}{\tau_X}}+R^2e^{-\frac{\Delta t}{\tau_Y}}+ 
      2R\,e^{-\frac{\Delta t (\tau_X + \tau_Y)}{2(\tau_X\,\tau_Y)}} \cos\left(\delta\, \Delta t+\phi\right)\right].
\end{eqnarray}

Using the measured coherence times $\tau_X$ and  $\tau_Y$ and the values of
$R$ and $\phi$ calculated from the interaction strengths of the transitions,
we fit the experimental data in figure~\ref{fig:interference_beats} using
(\ref{eq:model}) with only $G_0$ and an accidental count rate as free
parameters, and find a good agreement with this relatively simple model.

Different interaction strengths for the different polarizations in the
participating levels allow control of the relative amplitude and phase of the
possible decay paths.
We can observe the dependence of the amplitude of the oscillation on the polarization settings and compare it with the expected values calculated from the interaction strengths of the transitions.
In figure~\ref{fig:beats_phase} we present three different cases, all fitted in a similar way as for figure~\ref{fig:interference_beats}.
Of particular interest is the case where the beats are almost entirely
suppressed [figure~\ref{fig:beats_phase}(a)], an indication that most of the
photon pairs observed are generated by the $X$ decay. Figure
~\ref{fig:beats_phase}(b) and (c) show the situation for polarization
selections that lead to quantum beats with a high contrast, but opposite
phases.

This demonstrates that it is possible to select one frequency component only
by an appropriate choice of polarizations, without using an etalon.
However, it is 
not possible to select only photon pairs from the $Y$ decay in a similar
manner due to the relative weakness of the transitions involving the
$5$P$_{3/2},\,F=2$ level.

\section{Conclusion}
In summary, we have characterized the polarization entangled state of photon
pairs from a cold cloud of atoms by performing quantum state
tomography, individually for two decay paths of the cascade. We find that the
resulting polarization-entangled states for both decay paths are not maximally
entangled, but reasonably close to it. 
This is compatible with a  model combining the transition strengths
between different participating intermediate states in the four-wave mixing
process.
We observe  high-contrast quantum beats in a time correlation measurement between the generated photon pairs. The contrast and the initial phase of beats can be controlled  with the choice of polarization of pumps and projective measurements on the generated
photons.

\ack{We acknowledge the support of this work by the National Research Foundation
(partly under grant No. NRF-CRP12-2013-03) \& Ministry of Education in
Singapore.}

\bibliographystyle{iopart-num}
\providecommand{\newblock}{}

\end{document}